\documentstyle[twocolumn,aps]{revtex}

\begin{document}
\title{Gauge Supergravities for all Odd Dimensions\thanks{Talk presented at the Second Meeting Quantum Gravity in the Southern Cone, Bariloche, Argentina, January 1998.}}
\author{Ricardo Troncoso and Jorge Zanelli\thanks{John Simon Guggenheim fellow}}
\address{Centro de Estudios Cient\'{\i}ficos de Santiago, Casilla 16443, Santiago 9, Chile\\ and \\
Departamento de F\'{\i}sica, Universidad de Santiago de Chile, Casilla 307, Santiago 2, Chile.}
\maketitle

\begin{abstract}
Recently proposed supergravity theories in odd dimensions whose fields are connection one-forms for the minimal supersymmetric extensions of anti-de Sitter gravity are discussed.  Two essential ingredients are required for this construction: (1) The superalgebras, which extend the adS algebra for different dimensions, and (2) the lagrangians, which are Chern-Simons $(2n-1)$-forms.  The first item  completes the analysis of van Holten and Van Proeyen, which was valid for $N=1$ only. The second ensures that the actions are invariant by construction under the gauge supergroup and, in particular, under local supersymmetry. Thus, unlike standard supergravity, the local supersymmetry algebra closes off-shell and without requiring auxiliary fields.\\
 
The superalgebras are constructed for all dimensions and they fall into three families: $osp(m|N)$ for $D=2,3,4$, mod 8, $osp(N|m)$ for $D=6,7,8$, mod 8,  and $su(m-2,2|N)$ for $D=5$ mod 4, with $m=2^{[D/2]}$. The lagrangian is constructed for $D=5,\;7$ and 11. In all cases the field content includes the vielbein ($e_{\mu }^{a}$), the spin connection ($\omega _{\mu }^{ab}$), $N$ gravitini ($\psi _{\mu }^{i}$), and some extra bosonic ``matter" fields which vary from one dimension to another.  
\end{abstract}

\preprint{\begin{tabular}{l} hep-th/9807XXX\end{tabular}}

\section{Introduction}

This is an expanded version of a recent paper \cite{trz} and the lecture \cite{trz'}, where most of the preliminary results were announced.

Three of the four fundamental forces of nature are consistently described by Yang-Mills ({\bf YM}) quantum theories. Gravity, the fourth fundamental interaction, resists quantization in spite of several decades of intensive research in this direction. This is intriguing in view of the fact that General Relativity ({\bf GR}) and YM theories have a deep geometrical nature based on the gauge principle. How come two theories constructed on almost the same mathematical foundation produce such radically different physical
behaviours? What is the obstruction for the application of the methods of YM quantum field theory to gravity?  The final answer to these questions is beyond the scope of this paper, however one can note a difference between YM and GR which might turn out to be an important clue: YM theory is defined on a fiber bundle, with the connection as the dynamical object, whereas the dynamical fields of GR cannot be interpreted as components of a connection.

The closest one could get to a connection formulation for GR is the Palatini formalism, with the Hilbert action 
\begin{equation}
I[\omega ,e]=\int \epsilon _{abcd}R^{ab}\wedge e^{a}\wedge e^{b},
\label{hilbert}
\end{equation}
where $R^{ab}=d\omega ^{ab}+\omega _{c}^{a}\wedge \omega _{b}^{c}$ is the
curvature two-form. This action is sometimes --mistakenly-- claimed to describe a gauge theory for local translations.  If $\omega$ and $e$ were the components of the Poincar\'{e} connection associated to local translations, they should transform as
\begin{equation}
\delta \omega ^{ab}=0,\;\;\delta e^{a}=D\lambda ^{a}=d\lambda ^{a}+\omega
_{b}^{a}\wedge \lambda ^{b}.  \label{trans}
\end{equation}
Invariance of (\ref{hilbert}) under (\ref{trans})would require the torsion-free condition, 
\begin{equation}
T^{a}=de^{a}+\omega _{b}^{a}\wedge e^{b}=0,
\label{T}
\end{equation}
which is an equation of motion for the action (\ref{hilbert}). This means
that the invariance of the action (\ref{hilbert}) under (\ref{diffeos})
does not result from the transformation properties of the fields alone, but it is a property of their dynamics as well.

The error stems from the identification between local translations in the
base manifold (diffeomorphisms) 
\begin{equation}
x^{\mu }\rightarrow x^{\prime }{}^{\mu }=x^{\mu }+\zeta ^{\mu }(x),
\label{diffeos}
\end{equation}
--which is a genuine invariance of the action (\ref{hilbert})--, and local translations in the tangent space (\ref{trans}).  

The torsion-free condition, being one of the field equations, implies that local translational invariance is at best an {\em on-shell} symmetry, which would probably not survive quantization.

Since the invariance of the Hilbert action under general coordinate transformations (\ref{diffeos}) is reflected in the closure of the first-class hamiltonian constraints in the Dirac formalism, one could try to push the analogy between the Hamiltonian constraints $H_{\mu}$ and the generators of a gauge algebra. However, the fact that the constraint algebra requires structure {\em functions}, which depend on the dynamical fields, is another indication that the generators of diffeomorphism invariance of the theory do not form a Lie algebra but an open algebra (see, e. g., \cite{MH}).

More precisely, the subalgebra of spatial diffeomorphisms {\em is} a genuine Lie algebra in the sense that its structure constants are independent of the dynamical fields of gravitation, 
\begin{equation}
[H_i, H_j^{\prime}] \sim H_j^{\prime}\delta_{|i} - H_i^{\prime}\delta_{|j},
\label{hi}
\end{equation}
whereas, the generators of timelike diffeomorphisms are the offending ones: they form an open algebra, 
\begin{equation}
[H_{\perp}, H_{\perp}^{\prime}] \sim g^{ij}H_j^{\prime}\delta_{|i}.
\label{hperp}
\end{equation}

This comment is particularly appropriate in a CS theory, where spatial
diffeomorphisms are always part of the true gauge symmetries of the theory. The generators of timelike displacements ($H_{\perp}$), on the other hand, are combinations of the internal gauge generators and the generators of spatial diffeomorphism, and therefore do not generate independent symmetries \cite{bgh}.

\section{Supergravity}

For some time it was hoped that the nonrenormalizability of GR could be
cured by supersymmetry. However, the initial glamour of supergravity ({\bf SUGRA}) as a mechanism for taming the wild ultraviolet divergences of pure gravity, was eventually spoiled by the realization that it too would lead to a nonrenormalizable answer. Again, one can see that SUGRA is not a gauge theory in the sense of a fiber bundle and that the local symmetry algebra closes naturally only on shell. The algebra can be made to close off shell at the cost of introducing auxiliary fields, but they are not guaranteed to exist for all $D$ and $N$ \cite{tr}.

Whether the lack of fiber bundle structure is the ultimate reason for the
nonrenormalizability of gravity remains to be proven. However, it is
certainly true that if GR could be formulated as a gauge theory, the chances of proving its renormalizability would clearly grow. In three spacetime dimensions both GR and SUGRA define renormalizable quantum theories. It is strongly suggestive that precisely in this case both theories can also be formulated as gauge theories on a fiber bundle \cite{witten}. It might seem that the exact solvability miracle was due to the absence of propagating degrees of freedom in three-dimensional gravity, but the key ingredient of the miracle can be traced down to the fiber bundle structure of the Chern-Simons ({\bf CS}) form of those systems. 

There are other known examples of gravitation theories in odd dimensions which are genuine (off-shell) gauge theories for the anti-de Sitter ({\bf adS}) or Poincar\'{e} groups \cite{chamslett,chamseddine,btz,z}. These theories, as well as their supersymmetric extensions have propagating degrees of freedom \cite{bgh} and are CS systems for the corresponding groups as shown in \cite{btrz}.

\subsection{From Rigid Supersymmetry to Supergravity}

Rigid SUSY can be understood as an extension of the Poincar\'e algebra by including supercharges which are the ``square roots" of the generators of rigid translations, $\{\bar{Q},Q\} \sim \Gamma\cdot \mbox{P}$. Roughly speaking, the traditional strategy to generalize this idea to local SUSY was to substitute the momentum P$_{\mu} = i\partial_{\mu}$ by the generators of diffeomorphisms, $\{\bar{Q},Q\} \sim \Gamma\cdot{\cal H}$. The resulting theory has an on-shell local supersymmetry algebra.

An alternative point of view --which is the one we advocate here-- is to
construct the supersymmetry on the tangent space and not on the base
manifold. This approach is more natural if one recalls that spinors provide a basis of irreducible representations for $SO(N)$, and not for $GL(N)$. Thus, spinors are naturally defined relative to a local frame on the tangent space rather than in the coordinate basis. This idea has been successfully applied by Chamseddine in five dimensions\cite{chamseddine}, and by us for pure gravity \cite{btz,z} and in supergravity \cite{trz,btrz}. The basic point is to replicate the 2+1 ``miracle" in higher dimensions. The construction has been carried out for spacetimes whose tangent space has adS symmetry in \cite{trz}, and for its Poincar\'e contraction in \cite{btrz}.

In \cite{btrz}, a family of theories in odd dimensions, invariant under the
supertranslation algebra whose bosonic sector contains the Poincar\'{e}
generators was presented. The anticommutator of the supersymmetry generators
gives a translation plus a tensorial ``central" extension, 
\begin{equation}
\{Q^{\alpha },\bar{Q}_{\beta }\}=-i(\Gamma ^{a})_{\beta }^{\alpha}P_{a}
-i(\Gamma^{abcde})_{\beta }^{\alpha }Z_{abcde},  \label{supertrans}
\end{equation}
The commutators of $Q,\bar{Q}$ and $Z$ with the Lorentz generators can be
read off from their tensorial character. All the remaining commutators
vanish. This algebra is the continuation to all odd-dimensional spacetimes
of the $D=10$ superalgebra of Ref. \cite{vV} and yields supersymmetric
theories with off-shell Poincar\'{e} superalgebra. The existence of these
theories suggests that there should be similar supergravities based on the
adS algebra.

\subsection{Assumptions of Standard Supergravity}

Three implicit assumptions are usually made in the construction of standard SUGRA:

{\bf (i)} The fermionic and bosonic fields in the lagrangian should come in combinations such that their propagating degrees of freedom are equal in number.  This is usually achieved by adding to the graviton and the gravitini a number of lower spin fields ($s<3/2$)\cite{pvn}. This matching, however, is not necessarily true in adS space, nor in Minkowski space if a different representation of the Poincar\'e group (e.g., the adjoint representation) is used \cite{soh}.

The other two assumptions concern the purely gravitational sector. They are as old as General Relativity itself and are dictated by economy: {\bf (ii)} gravitons are described by the Hilbert action (plus a possible
cosmological constant), and, {\bf (iii)} the spin connection and the vielbein are not independent fields but are related through the torsion
equation. The fact that the supergravity generators do not form a closed
off-shell algebra can be traced back to these asumptions.

The procedure behind {\bf (i)} is tightly linked to the idea that the fields should be in a {\em vector} representation of the Poincar\'{e} group \cite{soh} and that the kinetic terms and couplings are such that the counting of degrees of freedom works like in a minimally coupled gauge theory. This assumption comes from the interpretation of supersymmetric
states as represented by the in- and out- plane waves in an asymptotically
free, weakly interacting theory in a minkowskian background. These
conditions are not necessarily met for a CS theory in an asymptotically adS background. Apart from the difference in background, which requires a
careful treatment of the unitary irreducible representations of the
asymptotic symmetries \cite{gs}, the counting of degrees of freedom in CS
theories is completely different from the one for the same connection
one-forms in a YM theory.

\subsection{Lanczos--Lovelock Gravity}

For $D>4$, assumption {\bf (ii)} is an unnecessary restriction in the
available theories of gravitation. In fact, the most general action for
gravity --generally covariant and with second order field equations for the metric-- is a polynomial of degree $[D/2]$ in the curvature, first discussed by Lanczos \cite{lanczos} for $D=5$ and, in general, by Lovelock \cite{lovelock,zumino}. This action contains the same degrees of freedom as the Hilbert action \cite{tz} and is the most general low-energy effective theory of gravity derived from string theory \cite{zwiebach}.

The Lanczos-Lovelock ({\bf LL}) theory contains as a particular case the Einstein-Hilbert ({\bf EH}) theory but they are in general dynamically quite different. The classical solutions of the LL theory are not perturbatively related to those of Einstein's theory. For instance, it was observed that the time evolution of the classical solutions in the LL theory starting from a generic initial state can be unpredictable, whereas the EH theory defines a well-posed Cauchy problem \cite{tz,htz}. Moreover, the LL theory has a large number of dimensionful constants in contrast with the two constants of the EH action ($g$ and $\Lambda$) \cite{bd,jjg,btz}). This last feature would seem to indicate that renormalizability would be even more remote for the LL theory than in ordinary gravity.

However, there is an exceptional case when all the constants of the LL action are related to each other in a particular way [c.f., next section]. The resulting system exhibits a larger symmetry, the theory can be formulated in terms of a dimensionless connection 1-form and a unique dimensionless constant which is also quantized \cite{z}. In this case, the LL theory becomes a CS system in which the spin connection and the vielbein are parts of an adS connection. A particular case of this occurs naturally in 2+1 dimensions in the presence of a cosmological constant. This is the secret behind the integrability of gravity in 2+1 dimensions as demonstrated in \cite{witten}.

\subsection{Torsion}

Assumption {\bf (iii)} implies that torsion does not contain independently
propagating degrees of freedom, and its equations are identities enforced by fiat on the fields. This is the essence of the metric approach to GR in
which parallel transport and distance are not independent notions but are
related through the Christoffel symbols. There is no natural justification for this assumption and it was the leit motif of the historic discussion between Einstein and Cartan \cite{E-C}.

In four dimensions if $e$ and $\omega$ are varied independently in the
action (\ref{hilbert}), the equation for $\omega$ implies $T^a=0$ (in the absence of matter). This might seem sufficient to justify asumption {\bf (iii)} since the equation for $\omega$ is algebraic and could in principle be used to express $\omega$ in terms of the remaining fields. However, it is not always possible to solve this algebraic equation for $\omega$ if $D>4$.

Another serious consequence of the torsion-free conditions is that the dynamical dependence between $\omega $ and $e$ introduced by (\ref{T}) spoils the possibility of interpreting the local translational invariance as a gauge symmetry of the action. Indeed, taking the condition $T^{a}=0$ as a definition, would imply
\begin{equation}
\delta \omega ^{ab}= \frac{\delta \omega ^{ab}}{\delta e^{c}}\delta e^{c} \neq 0,
\label{deltae}
\end{equation}
in contrast with (\ref{trans}).  Thus, the spin connection and the vielbein cannot be identified as the compensating fields for local Lorentz rotations and translations, respectively, as it can be done in $D=3$. Thus, General Relativity in $D=2n\geq 4$ cannot be formulated purely as a gauge theory on a fiber bundle.

In our construction \cite{trz,btrz}) $\omega$ and $e$ are assumed to be dynamically independent and thus torsion necessarily contains propagating degrees of freedom, represented by the contorsion tensor $k^{ab}_{\mu}:= \omega^{ab}_{\mu} - \bar{\omega}^{ab}_{\mu}(e,...)$, where $\bar{\omega}$ is the spin connection which solves the classical torsion equation in terms of the remaining fields.

\section{Gauge Gravity and Gauge Supergravity}

\subsection{Chern-Simons Gravity in $2n-1$ Dimensions}

For $D=2n-1$, the particular choice of the LL lagrangian reads 
\begin{equation}
L_{G\;2n-1}^{adS}=\sum_{p=0}^{n-1}\alpha _{p}L^{p},  \label{LL}
\end{equation}
where 
\begin{equation}
L_{G}^{p}=\epsilon _{a_{1}\cdots a_{D}}R^{a_{1}a_{2}}\cdots
R^{a_{2p-1}a_{2p}}e^{a_{2p+1}}\cdots e^{a_{D}}.  \label{L-adS}
\end{equation}
Here wedge product is understood and the subscript ``$G$'' stands for
``torsion-free gravity''. We note that although torsion doesn't explicitly enter in (\ref{L-adS}), for $D\geq 5$ this system has propagating torsion. 

If one chooses $\alpha _{p}=\kappa (D-2p)^{-1}(_{\;\;p}^{n-1})l^{2p-D}$ the invariance of (\ref{LL}) under the Lorentz group extends to the adS group. The constant $l$ has dimensions of length and its purpose is to render the action dimensionless allowing the interpretation of $\omega $ and $e$ as components of the adS connection \cite{jjg} 
\begin{equation}
W^{AB}=\left[ \begin{array}{cc}
\omega ^{ab} & e^{a}/l \\ 
-e^{b}/l & 0
\end{array} \right] ,  
\label{W}
\end{equation}
where $A,B=1,...D+1$. The resulting lagrangian is an adS-CS form in the sense that its exterior derivative is the Euler form in $2n$ dimensions \cite{chamseddine,btz,jjg}, 
\begin{equation}
dL_{G\;2n-1}^{adS}=\kappa \epsilon _{A_{1}\cdots A_{2n}}R^{A_{1}A_{2}}\cdots
R^{A_{2n-1}A_{2n}},  \label{E}
\end{equation}
where $R^{AB}$ is the adS curvature and $\kappa $ is quantized \cite{z} (in the following we will set $\kappa =l=1$). 

Apart from the Euler-CS form, discussed above, there are the standard CS ($2p-1$)-forms $L_{2p-1}^{*}$, related to the $p^{th}$ Chern character for the Lorentz connection,
\begin{equation}
dL_{2p-1}^{*}(\omega _{b}^{a})=c_{p},  \label{Ccharacter}
\end{equation}
where $c_{p}=:Tr[(R_{b}^{a})^{p}]$.

In general, a Chern-Simons D-form is defined by the condition that its
exterior derivative be an invariant homogeneous polynomial of degree $n$ in the curvature, that is, a characteristic class. In the examples above, (\ref{E}) is the CS form for the Euler class $2n$-form, while the exotic lagrangians are related to different combinations of Chern characters [see Appendix A].

Thus, a generic CS action in $2n-1$ dimensions for a Lie algebra $g$ can be written as 
\begin{equation}
dL_{2n-1}^{g}=\left\langle\mbox{{\bf F}}^{n}\right\rangle ,  \label{F^n}
\end{equation}
where $\left\langle \ \right\rangle $ stands for a multilinear function in the Lie algebra $g$,
invariant under cyclic permutations such as Tr, for an ordinary Lie algebra, or STr, in the case of a superalgebra. For $D=3$ the ``exotic gravity" lagrangian 
\begin{equation}
L^{adS}_{T\;3}= \omega^a_b d\omega^b_a + \frac{2}{3} \omega^a_b \omega^b_c
\omega^c_a +2e_a T^a,  \label{L*3}
\end{equation}
is a CS form for the adS group$\footnote{For simplicity we will not distinguish between different signatures. Thus, the adS algebra in $D$ dimensions will be denoted as $so(D+1)$.} $SO(4), whose exterior derivative is the Pontryagin form in 4 dimensions ($2^{nd}$ Chern character for $SO(4)$) \cite{witten}, 
\begin{equation}
dL^{adS}_{T\;3}= R^A_B R^B_A,  \label{exotic}
\end{equation}
where $R^A_B$ is the Riemann curvature 2-form in four dimensions.

Similar exotic actions associated to the Chern characters in $4k$ dimensions exist in $D=4k-1$ \cite{tronco}. The gravitational Chern characters vanish for $p=2n+1$ (the trace of product of an odd number of Riemann curvatures vanishes) \cite{mz}, hence there are no adS CS lagrangians in $D=4k+1$. For $D=4k-1$,  the number of possible exotic forms grows as the partitions of $k$, in correspondence with the number of composite Chern invariants of the form $\prod_{i}c_{p_{i}}$, with $\sum_{i}p_{i}=4k$. Out of all these forms, we will be interested in particular combinations of them which, in the spinorial representation of $SO(4k)$, can be written as \cite{trz} 
\begin{equation}
dL_{T\;4k-1}^{adS}=-Tr[(\frac{1}{4}R^{AB}\Gamma _{AB})^{2k}].  \label{LT}
\end{equation}

It is important to note that in this lagrangian, as well as in (\ref{exotic}), torsion appears explicitly. For example, in seven dimensions one finds 
\begin{eqnarray*}
&&L_{T\;7}^{adS}= L_7^*(\omega )-\!\frac{3}{4}(R^a\,_bR^b\,_a + 2[T^aT_a
\!-\! R^{ab}e_a e_b]) L^*_3(\omega)  \nonumber \\
&&-(\frac{3}{2}R^a\,_bR^b\,_a + 2 T^aT_a -\! 4R^{ab}e_a e_b)T^a e_a + 4T_a
R^a\,_bR^b\,_a e^c.  
\end{eqnarray*}
In eight dimensions there are 4 topological invariants \cite{cz}, which give rise to four corresponding CS forms in seven dimensions.  $L_{T\;7}^{adS}$ is a particular linear combination of those CS forms. For further discussion on these invariants, see Ref. \cite{tronco,cz}.

If one allows the presence of torsion explicitly the maximal extension of the LL lagrangian is found.  This is the most general $D$-form invariant under local Lorentz rotations constructed out of the vielbein, the spin connection and their derivatives (without using the metric) \cite{mz}. Lagrangian (\ref{LT}) is a particular representative of this family in which the coefficients are chosen so as to make the action locally invariant under the adS group $SO(D+1)$ \cite{tronco}.

The exterior derivative af a possible CS lagrangian for a given group in $D=2n-1$ dimensions has the form $g_{a_{1}\cdots a_{n}}F^{a_{1}}\cdots
F^{a_{n}}$, where $g_{a_{1}\cdots a_{n}}$ is an invariant tensor of the algebra. Thus, the  problem of finding all possible CS lagrangians is equivalent to finding all possible invariant tensors of rank $n$ in the Lie algebra. This is in general an open problem, related to the number of Casimir invariants for a given Lie group.  For the groups relevant for supergravity discussed below, like $OSp(32|1)$, the number of invariant tensors can be rather large. Most of these invariants would give rise to bizarre lagrangians and the real problem is to find the appropriate invariants that describe sensible theories.

The R.H.S. of (\ref{LT}) is a particular form of (\ref{F^n}) where $\left\langle \ \right\rangle $ is the ordinary trace over spinor indices. Other possibilities of the form $\left\langle \mbox{{\bf F}}^{n-p} \right\rangle \left\langle\mbox{{\bf F}}^p \right\rangle $, are not used in our construction as they would not lead to the minimal supersymmetric extensions of adS containing the Hilbert action. In the supergravity theories discussed below, the gravitational sector is given by $\pm\frac{1}{2^n} L_{G\;2n-1}^{adS}- \frac{1}{2}L_{T\;2n-1}^{adS}$. The $\pm $ sign corresponds to the two choices of inequivalent representations of $\Gamma $'s, which in turn reflect the two chiral representations in $D+1$. As in the three-dimensional case, the supersymmetric extensions of $L_{G}$ or any of the exotic lagrangians such as $L_{T}$, require using both chiralities, thus doubling the algebras. Here we choose the + sign, which gives the minimal superextension \cite{tronco}. 

The bosonic theory outlined above is our starting point. The idea now is to construct its supersymmetric extension.  A possible approach would be to study the superalgebras containing anti-de Sitter as a subalgebra, define a connection --in the adjoint representation-- and construct the CS form with it. This construction must clearly give the right result, but it would be rather difficult to proceed formally without an explicit representation. Knowing this, we take a less formal path, expressing the adjoint representation in terms of the Dirac matrices of the appropriate dimension. This is not a wild guess because the generators of the Dirac algebra, $\{I$, $\Gamma ^{a}$, $\Gamma ^{ab}$,...$\}$, is a basis for all square matrices. The advantage of this approach is that it provides an explicit representation of the algebra and writing the lagrangians is straightforward.\\

\subsection{Gauge Supergravity}

The supersymmetric extensions of the adS algebras in $D=$ 2, 3, 4, mod 8, were studied by van Holten and Van Proeyen in \cite{vV}. They added one Majorana supersymmetry generator to the adS algebra and found all the $N=1$ extensions demanding closure of the full superalgebra. In spite of the fact that the algebra for $N=1$ adS supergravity in eleven dimensions was conjectured in 1978 to be $osp(32|1)$ by Cremer, Julia and Scherk \cite{CJS}, and this was confirmed in \cite{vV}, nobody constructed a supergravity action for this algebra in the intervening twenty years. One reason for the lack of interest in the problem might have been the fact that the $osp(32|1)$ algebra contains generators which are Lorentz tensors of rank higher than two.

Apart from the assumptions mentioned above, supergravity algebras were traditionally limited to generators which are Lorentz tensors up to second rank.  This constraint was based on the observation that elementary particle states of spin higher than two would be inconsistent \cite{nahm}. However, this does not rule out the relevance of those tensor generators in theories of extended objects \cite{AGIT}. In fact, it is quite common nowadays to find algebras like the $M-$brane superalgebra \cite{townsend,NG}, 
\begin{equation}
\{Q^{,}\bar{Q}\}\sim \Gamma ^{a}P_{a}+ \Gamma ^{ab}Z_{ab}+ \Gamma^{abcde} Z_{abcde}.
\label{malgebra}
\end{equation}

\section{Superalgebra and Connection}

The smallest superalgebra containing the adS algebra in the bosonic sector
is found following the same approach as in \cite{vV}, but lifting the
restriction of $N=1$ \cite{tronco}. The result, for odd $D>3$ is (see Appendix B) \newline

\begin{center}
\begin{tabular}{|l|c|c|c|}
\hline
D & S-Algebra & Conjugation Matrix & Internal Metric \\ \hline
$8k-1$ & $osp(N|m)$ & $C^{T}=C$ & $u^{T}=-u$ \\ \hline
$8k+3$ & $osp(m|N)$ & $C^{T}=-C$ & $u^{T}=u$ \\ \hline
$4k+1$ & $su(m|N)$ & $C^{\dag }=C$ & $u^{\dag }=u$ \\ \hline
\end{tabular}
\end{center}

In each of these cases, $m=2^{[D/2]}$ and the connection takes the form 
\begin{eqnarray}
\mbox{{\bf A}}&=& \frac{1}{2}\omega ^{ab}J_{ab}+ e^{a}J_{a} + \frac{1}{r!}
b^{[r]}Z_{[r]}+  \nonumber \\
&&\frac{1}{2}(\bar{\psi}^i Q_i -\bar{Q}^i \psi_i) + \frac{1}{2}a_{ij}M^{ij}.
\label{A}
\end{eqnarray}

The generators $J_{ab},J_{a}$ span the adS algebra, $Q_{\alpha}^i$ generate
(extended) supersymmetry transformations, and $[r]$ denotes a set of $r$
antisymmetrized Lorentz indices. The $Q^{\prime}s$ transform as vectors
under the action of $M_{ij}$ and as spinors under the Lorentz group. Finally, the $Z$'s complete the extension of adS into the larger algebras $so(m)$, $sp(m)$ or $su(m)$.

In (\ref{A}) $\bar{\psi}^i=\psi_j^TCu^{ji}$ ($\bar{\psi}^i=\psi_j^{\dag} Cu^{ji}$ for $D=4k+1$), where $C$ and $u$ are given in the table above. These algebras admit $(m+N)\times (m+N)$ matrix representations \cite{freund}, where the $J$ and $Z$ have entries in the $m\times m$ block, the $M_{ij}$'s in the $N\times N$ block, while the fermionic generators $Q$ have entries in the complementary off-diagonal blocks.

Under a gauge transformation, {\bf A} transforms by $\delta${\bf A}$= \nabla\lambda$, where $\nabla$ is the covariant derivative for the same connection {\bf A}. In particular, under a supersymmetry transformation, $\lambda =\bar{\epsilon}^iQ_i-\bar{Q}^i\epsilon_i$, and 
\begin{equation}
\delta_{\epsilon}\mbox{{\bf A}}=\left[ \begin{array}{cc}
\epsilon^k\bar{\psi}_k- \psi^k\bar{\epsilon}_k & D\epsilon_j \\ 
-D\bar{\epsilon}^i & \bar{\epsilon}^i\psi_j-\bar{\psi}^i\epsilon_j \end{array} \right],  
\label{delA}
\end{equation}
where $D$ is the covariant derivative on the bosonic connection, $D\epsilon_j =(d+\frac{1}{2}[e^a\Gamma_a +\frac{1}{2}\omega^{ab}\Gamma_{ab}+ \frac{1}{r!} b^{[r]}\Gamma_{[r]}])\epsilon_j -a_j^i \epsilon_i$.

\subsection{D=5 Supergravity}

In this case, as in every dimension $D=4k+1$, there is no torsional Lagrangians $L_{T}$ due to the vanishing of the Pontrjagin $4k+2$-forms for the Riemann cirvature. This fact implies that the local supersymmetric extension will be of the form $L=L_{G}+\cdot \cdot \cdot $.

As shown in the previous table, the appropriate $adS$ superalgebra in five
dimensions is $su(2,2|N),$ whose generators are $K,J_a,J_{ab},Q^{\alpha },
\bar{Q}_{\beta}, M^{ij}$, with $a,b=1,...,5$ and $i,j=1,...,N$. The connection is {\bf A}$=b K+e^a J_a +\frac{1}{2}\omega^{ab}J_{ab}+ a_{ij}M^{ij}+ \bar{\psi}^iQ_i -\bar{Q}^j\psi_j$, so that in the adjoint representation 
\begin{equation}
\mbox{{\bf A}}=\left[ 
\begin{array}{cc}
\Omega_{\beta}^{\alpha} & \psi_j^{\alpha} \\ 
-\bar{\psi}_{\beta}^i & A_j^i
\end{array} \right],
\label{cone5}
\end{equation}
with $\Omega_{\beta}^{\alpha}=\frac{1}{2}(\frac{i}{2}bI+ e^a \Gamma_a+ \omega^{ab} \Gamma_{ab})_{\beta}^{\alpha}$, $A_j^i=\frac{i}{N}\delta_j^ib+ a_j^i$, and $\bar{\psi}_{\beta}^i=\psi^{\dagger \alpha j}G_{\alpha \beta}$. Here $G$ is the Dirac conjugate (e. g., $G=i\Gamma^0$). The curvature is
\begin{equation}
\mbox{{\bf F}}=\left[ \begin{array}{cc}
\bar{R}_{\beta}^{\alpha} & D\psi_j^{\alpha} \\ 
-D\bar{\psi}_{\beta}^i & \bar{F}_j^i
\end{array} \right]  
\label{curva5}
\end{equation}
where
\begin{eqnarray}
D\psi_j^{\alpha}&=&d\psi_j^{\alpha}+ \Omega_{\beta}^{\alpha}\psi_j^{\beta}- A_j^i\psi_i^{\alpha},  \nonumber \\
\bar{R}_{\beta}^{\alpha}&=&R_{\beta}^{\alpha}-
\psi_i^{\alpha}\bar{\psi}_{\beta}^i, \\
\bar{F}_j^i&=&F_j^i-\bar{\psi}_{\beta}^i\psi_j^{\beta}. 
\nonumber
\end{eqnarray}
Here $F_j^i=dA_j^i + A_k^i A_j^k +\frac{i}{N}db \delta^i_j$ is the $su(N)$ curvature, and $R_{\beta}^{\alpha} = d\Omega_{\beta}^{\alpha}+ \Omega_{\sigma}^{\alpha} \Omega_{\beta}^{\sigma}$ is the $u(2,2)$ curvature.  In terms of the standard $(2n-1)$-dimensional fields, $R_{\beta}^{\alpha}$ can be written as
\begin{equation}
R^{\alpha}_{\beta}=\frac{i}{4}db \delta^{\alpha}_{\beta}+ \frac{1}{2}\left[T^a\Gamma_a+ (R^{ab}+ e^ae^b)\Gamma_{ab} \right]^{\alpha}_{\beta}.
\end{equation}

In six dimensions the only invariant form is

\begin{equation}
P=iStr\left[\mbox{{\bf F}}^3 \right] ,
\end{equation}
which in this case reads
\begin{eqnarray}
P &=&Tr\left[R^3\right] -Tr\left[F^3\right] \\
&+&3\left[D\bar{\psi}(\bar{R}+ \bar{F})D\psi -\bar{\psi}(R^2-F^2 + [R-F](\psi)^2 )\psi \right] ,  \nonumber
\end{eqnarray}
where $(\psi)^2 =\bar{\psi}\psi$. The resulting five-dimensional C-S density can de descompossed as a sum a a gravitational part, a $b$-dependent piece,  a $su(N)$ gauge part, and a fermionic term,
\begin{equation}
L=L_{G}^{adS}+L_{b}+L_{su(N)}+L_{F},
\end{equation}
with
\begin{equation}
\begin{array}{l}
L_G^{adS}= \frac{1}{8}\epsilon_{abcde}(R^{ab}R^{cd}e^e+ \frac{2}{3}
R^{ab}e^ce^de^e+ \frac{1}{5}e^ae^be^ce^de^e) \\ 
 \\
L_b = -(\frac{1}{N^2}-\frac{1}{4^2})(db)^2b+\frac{3}{4}(T^aT_a -R^{ab}e_a e_b - \frac{1}{2}R^{ab}R_{ab})b +\frac{3}{N}b f^i_jf^j_i\\ 
 \\
L_{su(N)} = -(a_j^i da_k^j da_i^k+ a_j^i a_k^j a_l^k da_i^l+ \frac{3}{5} a_j^ia_k^j a_l^k a_m^l a_i^m)  \\ 
 \\
L_{F} = \frac{3}{2}\left[\bar{\psi}(\bar{R}+\bar{F})D\psi- \frac{1}{2}(\psi)^2(\bar{\psi}D\psi)\right]. 
\end{array}
\end{equation}

The action is invariant under local gauge transformations, which contain the local SUSY transformations
\begin{equation}
\begin{array}{ccl}
\delta e^a &=&-\frac{1}{2}(\overline{\epsilon}^i \Gamma^a \psi_i- \overline{\psi}^i \Gamma^a \epsilon_i) \\ 
\delta \omega ^{ab} &=& \frac{1}{4}(\overline{\epsilon}^i \Gamma^{ab} \psi_i- \overline{\psi}^i \Gamma^{ab} \epsilon_i) \\ 
\delta b &=& i(\overline{\epsilon}^i\psi_i- \overline{\psi}^i\epsilon_i) \\ 
\delta \psi_i &=& D\epsilon_i \\ 
\delta \overline{\psi}^i &=& D\overline{\epsilon}^i \\ 
\delta a_j^i &=&i(\overline{\epsilon}^i \psi_j- \overline{\psi}^i \epsilon_j).
\end{array}
\label{tr5}
\end{equation}

As in $2+1$ dimensions, the Poincar\'{e} supergravity theory is recovered contracting the super $adS$ group.  Consider the following rescaling of the fields
\begin{equation}
\begin{array}{ccc}
e^a & \rightarrow & \frac{1}{\alpha}e^a \\ 
\omega^{ab} & \rightarrow & \omega^{ab} \\ 
b & \rightarrow & \frac{1}{3\alpha}b \\ 
\psi_i & \rightarrow & \frac{1}{\sqrt{\alpha}}\psi_i \\ 
\overline{\psi}^i & \rightarrow & \frac{1}{\sqrt{\alpha}}\overline{\psi}^i \\ 
a_j^i & \rightarrow & a_j^i.
\end{array}
\label{rescal5}
\end{equation}
Then, if the gravitational constant is also rescaled as $\kappa \rightarrow \alpha \kappa$, in the limit $\alpha \rightarrow \infty $ the action becomes that in \cite{btrz}, plus a $su(N)$ CS form,
\begin{eqnarray}
I &=& \frac{1}{8}\int[\epsilon_{abcde}R^{ab}R^{cd}e^e -R^{ab}R_{ab}b- \\ \nonumber
&& 2R^{ab}(\overline{\psi}^i\Gamma_{ab}D\psi_i +D\overline{\psi}^i \Gamma_{ab}\psi_i) + L_{su(N)}] .
\label{contraction}
\end{eqnarray}

The rescaling (\ref{rescal5}) induces a contraction of the super $adS$ algebra $su(m|N)$ into [super Poincar\'{e}]$\otimes su(N)$, where the second factor is an automorfism.

\subsection{D=7 Supergravity}

The smallest adS superalgebra in seven dimensions is $osp(2|8)$. The
connection (\ref{A}) is {\bf A} =$\frac{1}{2}\omega^{ab}J_{ab}+e^{a}J_{a}+ 
\bar{Q}^{i}\psi _{i}+\frac{1}{2}a_{ij}M^{ij}$, where $M^{ij}$ are the
generators of $sp(2)$. In the representation given above, the bracket $\left\langle \ \right\rangle $  is the supertrace and, in terms of the component fields appearing in the connection, the CS form is 
\begin{eqnarray}
L_7^{osp(2|8)}(\mbox{{\bf A}})&=& 2^{-4}L_{G\;7}^{adS}(\omega,e)-\frac{1}{2}
L_{T\; 7}^{adS}(\omega,e)  \nonumber \\
&& -L_7^{*sp(2)}(a)+L_{F}(\psi ,\omega ,e,a).
\end{eqnarray}
Here the fermionic Lagrangian is 
\begin{eqnarray*}
L_F &=&4\bar{\psi}^j(R^{2}\delta_j^i + Rf_j^i +(f^2)_j^i)D\psi_i \\ \nonumber
& &+4(\bar{\psi}^i \psi_j)[(\bar{\psi}^j\psi_k)(\bar{\psi}^k D\psi_i) - \bar{\psi}^j(R\delta_i^k + f_i^k)D\psi_k] \\
&&-2(\bar{\psi}^{i}D\psi _{j})[\bar{\psi}^{j}(R\delta
_{i}^{k}+f_{i}^{k})\psi _{k}+D\bar{\psi}^{j}D\psi _{i}],  \nonumber
\end{eqnarray*}
where $f_j^i= da_j^i+a_k^i a_j^k$, and $R=\frac{1}{4}(R^{ab} +e^a e^b)
\Gamma_{ab} + \frac{1}{2} T^a \Gamma_a$ are the $sp(2)$ and $so(8)$
curvatures, respectively. The supersymmetry transformations (\ref{delA})
read \\

\begin{tabular}{llll}
\hspace{1cm} & $\delta e^a =\frac{1}{2}\bar{\epsilon}^i\Gamma^a \psi_i $ & 
\hspace{1cm} & $\delta \omega ^{ab}=-\frac{1}{2}\bar{\epsilon}^i\Gamma^{ab}
\psi_i$ \\ 
 \\
\hspace{1cm} & $\delta \psi_i =D\epsilon_i$ & \hspace{1cm} & $\delta a_j^i =
\bar{\epsilon}^i \psi_j-\bar{\psi}^i \epsilon_j.$ \label{susy7}
\end{tabular}
 \\
 
Standard seven-dimensional supergravity is an $N=2$ theory (its maximal
extension is N=4), whose gravitational sector is given by the Einstein-Hilbert action with cosmological constant and with an $osp(2|8)$ invariant background\cite{D=7,Salam-Sezgin}. In the case presented here, the extension to larger $N$ is straighforward: the index $i$ is allowed to run from $2$ to $2s$, and the Lagrangian is a CS form for $osp(2s|8)$. 

\subsection{D=11 Supergravity}

In this case, the smallest adS superalgebra is $osp(32|1)$ and the
connection is {\bf A} =$\frac{1}{2}\omega^{ab}J_{ab} + e^aJ_a + \frac{1}{5!} b^{abcde}J_{abcde}+ \bar{Q}\psi$, where $b$ is a totally antisymmetric fifth-rank Lorentz tensor one-form. Now, in terms of the elementary bosonic and fermionic fields, the CS form in (\ref{F^n}) reads 
\begin{equation}
L_{11}^{osp(32|1)}(${\bf A}$)= L_{11}^{sp(32)}(\Omega )+L_{F}(\Omega,\psi),  \label{L11}
\end{equation}
where $\Omega\equiv \frac{1}{2}(e^{a}\Gamma_{a} + \frac{1}{2}
\omega^{ab}\Gamma_{ab}+ \frac{1}{5!}b^{abcde}\Gamma_{abcde})$ is an $sp(32)$ connection. The bosonic part of (\ref{L11}) can be written as 
\begin{eqnarray}
L_{11}^{sp(32)}(\Omega)&=&2^{-6} L_{G\;11}^{adS}(\omega,e) -\frac{1}{2}
L_{T\;11}^{adS}(\omega,e) + L_{11}^b(b,\omega,e).  \nonumber
\end{eqnarray}
The fermionic Lagrangian is 
\begin{eqnarray*}
L_{F} &=&6(\bar{\psi}R^{4}D\psi)-3\left[ (D\bar{\psi}D\psi )+(\bar{\psi}
R\psi)\right] (\bar{\psi}R^{2}D\psi)  \nonumber \\
& &-3\left[(\bar{\psi}R^3\psi)+(D\bar{\psi}R^2 D\psi)\right](\bar{\psi} D\psi)+ \\
& &2\left[ (D\bar{\psi}D\psi )^{2}+(\bar{\psi}R\psi )^{2}+(\bar{\psi}R\psi)
(D\bar{\psi}D\psi )\right] (\bar{\psi}D\psi),
\end{eqnarray*}
where $R=d\Omega +\Omega ^{2}$ is the $sp(32)$ curvature. The supersymmetry
transformations (\ref{delA}) read \newline
 \\
\begin{tabular}{llll}
\hspace{1cm} & $\delta e^{a}=\frac{1}{8}\bar{\epsilon}\Gamma ^{a}\psi$ & 
\hspace{1cm} & $\delta\omega^{ab}=-\frac{1}{8}\bar{\epsilon}\Gamma^{ab}\psi$ \\ 
 \\
\hspace{1cm} & $\delta \psi =D\epsilon $ & \hspace{1cm} & $\delta b^{abcde}= 
\frac{1}{8}\bar{\epsilon}\Gamma ^{abcde}\psi.$ \\ 
\\ \label{susy11}
\end{tabular}

Standard eleven-dimensional supergravity \cite{CJS} is an N=1 supersymmetric extension of Einstein-Hilbert gravity that cannot accomodate a cosmological constant \cite{B-D-H-S,D}. An $N>1$ extension of this theory is not known. 
In our case, the cosmological constant is necessarily nonzero by construction and the extension simply requires including an internal $so(N)$ gauge field coupled to the fermions, and the resulting lagrangian is an $osp(32|N)$ CS
form \cite{tronco}.

\section{Discussion}

The supergravities presented here have two distinctive features: The
fundamental field is always the connection {\bf A} and, in their simplest
form, these are pure CS systems (matter couplings are discussed below). As a
result, these theories possess a larger gravitational sector, including
propagating spin connection. Contrary to what one could expect, the
geometrical interpretation is quite clear, the field structure is simple
and, in contrast with the standard cases, the supersymmetry transformations
close off shell without auxiliary fields.

The field content compares with that of the standard supergravities in $D=5,7,11$ as follows:\newline

\begin{center}
\begin{tabular}{c|c|l|l|}
\cline{2-4}
\hspace{.7cm} & D & Standard supergravity & CS supergravity \\ \cline{2-4}
& 5 & $e^a_{\mu}$ $\psi^{\alpha}_{\mu}$  $\bar{\psi}_{\alpha \mu}$  & $e_{\mu}^a$ $\omega^{ab}_{\mu}$ $\psi^{\alpha}_{\mu}$ $\bar{\psi}_{\alpha \mu}$ $b$\\ \cline{2-4}
& 7 & $e^a_{\mu}$ $A_{[3]}$ $\psi^{\alpha i}_{\mu}$ $a^i_{\mu j}$ $\lambda^{\alpha}$ $\phi$ & $e_{\mu }^a$ $\omega ^{ab}_{\mu}$ $\psi_{\mu}^{\alpha i }$ $a_{\mu j}^i$ \\ \cline{2-4}
& 11 & $e^a_{\mu}$ $A_{[3]}$ $\psi^{\alpha}_{\mu}$ & $e_{\mu}^{a}$ $\omega
^{ab}_{\mu }$ $\psi_{\mu }^{\alpha }$ $b^{abcde}_{\mu }$ \\ \cline{2-4}
\end{tabular}
\end{center}

Some sector of these theories might be related to the standard supergravities if one identifies the totally antisymmetric part of the contorsion tensor in a coordinate basis, $k_{\mu \nu \lambda}$, with the abelian 3-form, $A_{[3]}$. In 11 dimensions one could also identify the antisymmetric part of $b$ with an abelian 6-form $A_{[6]}$, whose exterior derivative, $dA_{[6]}$, is the dual of $F_{[4]}=dA_{[3]}$. Hence, in $D=11$ the CS theory possibly contains the standard supergravity as well as some kind of dual version of it \cite{nishino,BBS}.

The field equations for these theories take the form
\begin{equation}
\left\langle\mbox{{\bf F}}^{n-1}G_A\right\rangle =0,
\label{eq}
\end{equation}
where $G_A$ are the generators of the superalgebra (see Appendix C). Spreading out these equations in terms of the Lorentz components ($\omega$, $e$, $b$, $a$, $\psi$) produce somewhat involved expressions. It is therefore reassuring to verify that in all these theories the anti-de Sitter space is a fully SUSY background, and that for $\psi=b=a=0$ there exist spherically symmetric, asymptotically adS standard black-hole solutions of the class discussed in \cite{jjg} as well as topological black holes of the type discussed in \cite{pmb}. As it will be discussed in a forthcoming paper, for the extreme cases, these solutions are also BPS states \cite{AOTZ}.

It would be natural to inquire about the stability or positivity of the
energy for the linearized excitations around these or other solutions. This
problem, however, is highly nontrivial. As shown in Ref. \cite{bgh}, the phase space of a C-S system splits up into separate regions where the symplectic form has different rank with radically different dynamical content.  Thus, each C-S action describes several distinct systems in fact. In a generic case (where the rank of the symplectic structure is maximal) the number of bosonic and fermionic degrees of freedom do not match --as it is also found in \cite{HIPT}--, so the system cannot be naively related to a standard SUGRA \cite{R-J}.

It is possible to incorporate matter into these theories through a minimal
coupling {\bf A}. This case has been recently discussed by Horava \cite{horava}. The matter currents must have equal-time commutators obeying the superalgebras above, which are typical for a system of extended objects. For $D=11$, for example, the matter content is that of a theory with (super-) 0-, 2- and 5--branes, whose respective worldhistories couple to the spin connection and the $b$ fields.

In the last few months a number of papers have appeared dealing with 11D C-S theories and $OSp(32|1)$ \cite{nishino,BBS,horava,r3,r4}. In \cite{horava,r4}, an interesting suggestion is given on how standard SUGRA could be obtained from a C-S theory of the type discussed here.
 \\
 
{\bf ACKNOWLEDGEMENTS} \\

The authors are grateful to R. Aros, M. Ba\~nados, O. Chand\'{\i}a, M. Contreras, A. Dabholkar, S. Deser, G. Gibbons, A. Gomberoff, M. G\"unaydin, M. Henneaux, C. Mart\'{\i}nez, F. M\'endez, S. Mukhi, R. Olea, C. Teitelboim and E. Witten for many enlightening discussions and helpful comments. We would also like to express our appreciation to the organizers of the meeting for the exciting atmosphere and warm hospitality in Bariloche.  This work was supported in part by grants 1960229, 1970151, 1980788 and 3960009 from FONDECYT (Chile), and 27-953/ZI-DICYT (USACH). Institutional support to CECS from Fuerza A\'{e}rea de Chile and a group of Chilean private companies (Business Design Associates, CGE, CODELCO, COPEC, Empresas CMPC, Minera Collahuasi, Minera Escondida, NOVAGAS and XEROX-Chile) is also acknowledged.

\section{Appendices}

\subsection{Chern-Simons Action for $D=2n-1$}

Consider the $(2n-1)$-dimensional CS lagrangian defined by Eq.(\ref{F^n}),
\begin{equation}
dL_{2n-1}^{g}=\left\langle\mbox{{\bf F}}^{n}\right\rangle.  
\label{F^n'}
\end{equation}

Integrating this equation, $L$ can be written as \cite{nakahara,E-G-H}

\begin{equation}
L=\frac{1}{(n+1)!}\int_0^1dt\left\langle\mbox{{\bf A}\tiny $\wedge$}(td\mbox{\bf A}+ t^2
\mbox{{\bf A}\tiny $\wedge$}{\bf A})^{n-1}\right\rangle +\alpha  \label{GCS}
\end{equation}
where $\alpha$ is an arbitrary closed $(2n-1)$-form ($d\alpha =0$). 

Under a gauge transformation defined by a group element $g(x)$, the connection transforms as

\begin{equation}
\mbox{{\bf A}}\rightarrow \mbox{{\bf A}}^{g}=g^{-1}\mbox{{\bf A}}g+g^{-1}dg.
\end{equation}

The corresponding change in the Chern-Simons form is 
\begin{equation}
L^{g}=L+d\beta +(-1)^{n-1}\frac{n!(n-1)!}{(2n-1)!}\left\langle
(g^{-1}dg)^{2n-1}\right\rangle ,  \label{lg}
\end{equation}
where the $2(n-1)$-form $\beta$ is a function of {\bf A}, and depends on $g$ through the combination $g^{-1}dg$. Thus, the action
\begin{equation}
I_{C-S}=\int_{\Sigma }L,  \label{ics}
\end{equation}
which describes a gauge theory for the group {\bf G}, changes under a finite gauge transformation as the integral of (\ref{lg}), where the second term is a boundary term, and the third is proportional to the winding number.

One can see from (\ref{F^n'}) that under an infinitessimal gauge transformation (connected to the identity), the variation of $L$ is a total derivative,
\begin{equation}
\delta _{\lambda}\mbox{{\bf A}}=\nabla \lambda ,  \label{CSL/sym}
\end{equation}
where $\lambda$ is an arbtrary algebra-valued zero-form and $\nabla \lambda =d\lambda + [\mbox{{\bf A}},\lambda]$. In fact, the R.H.S. of (\ref{F^n'}) is invariant under (\ref {CSL/sym}), e.g., $\delta L$ is an exact form.

\subsection{Extended Superalgebras}

Let us look for the smallest supersymmetric extension of the anti-de
Sitter algebra in dimension $D$. The strategy is as follows: The graded
algebra will necessarily be of the form 
\begin{equation}
\begin{array}{ccc}
\left[ B,B\right] & \sim & B, \\ 
\left[ B,F\right] & \sim & F, \\ 
\left\{ F,F\right\} & \sim & B,
\end{array}
\label{bigrad}
\end{equation}
where the bosonic subalgebra is assumed to contain the generators of the adS group. The fermionic generators, on the other hand, must be in a spin 1/2 representation of the Lorentz group. The question now is, what is the minimal number of additional generators that is necessary to close the algebra? As was shown by van Holten and Van Proeyen \cite{vV}, it is possible to construct the $N=1$ supersymmetric extensions of adS for $ D=2,3,4 $ mod 8, but not for the remaining dimensions. Here we show that by relaxing the $N=1$ condition it is possible to extend the result for the other cases.

In Ref. \cite{vV}, the result is found by imposing the Jacobi identity for the algebra (\ref{bigrad}). We will derive the same result and its extension by demanding that certain matrices constructed from Lorentz invariant tensors form a representation of the algebra in a trivial way. This procedure excludes the exceptional supergroups from our analysis, however there is only case in which an exceptional supergroup is related to adS, that is $F(4)$ for $(D=6)$ \cite{nahm}.

We will suppose that fermionic generators ($F$) are in the spin-1/2 representation of the Lorentz group,
\begin{equation}
F\sim Q^{\alpha }_i, \alpha =1,...,m; \;i=1,...,N,  
\label{F}
\end{equation}
where $m=2^{[D/2]}$, $N$ is the number of supersymmetric generators, and $i$ is an internal index unrelated to the Lorentz group.

In accordance with (\ref{bigrad}), we expect the generators of the adS group to be contained in the r.h.s. of the third anticommutator. A simple way to achieve this is by using the spinorial representation for the adS algebra
\begin{eqnarray}
J_{a} &=&\Gamma _{a} \\  \nonumber
J_{ab} &=&\frac{1}{2}\Gamma_{ab}.  
\label{J}
\end{eqnarray}

Additional constraints can be imposed on the spinors $Q$ in order to produce  smaller supersymmetric extensions of adS. There are only two restrictions compatible with Lorentz invariance (see, e.g., \cite{freund}): definite chirality (the $Q$'s are Weyl spinors), and ``reality'' (Majorana spinors).  Chirality is only defined for even $D$. Here we shall not require chirality, but we demand the spinors to satisfy the following reality condition (modified Majorana spinors)
\begin{equation}
\bar{\psi}^{\alpha}_i= C^{\alpha \beta}u_{ij}\psi^j_{\beta},  
\label{majorana}
\end{equation}
which can be imposed for any $D$, provided the spacetime signature is $s-t=0, 1, 2, 6, 7$ mod 8 \cite{freund}. Here $C$ and $u$ are invertible matrices, the charge conjugation matrix, and the invariant metric of the internal symmetry group, respectively. Their inverses are defined with raised and lowered indices, respectively,
\begin{eqnarray}
C_{\alpha \beta}C^{\beta \gamma} &=& \delta_{\alpha}^{\gamma} \\  \nonumber
u^{ij} u_{jk} &=& \delta^i_k.  
\label{invers}
\end{eqnarray}

Without loss of generality, these metrics can be assumed to possess definite symmetry properties
\begin{equation}
C_{\gamma \beta }=\lambda C_{\beta \gamma}, \;\;u^{lk}=\mu u^{kl},
\label{metrics}
\end{equation}
whith $\lambda, \mu =\pm 1$. The charge conjugation matrix is defined so that
\begin{equation}
(\Gamma^a)^T = \eta C \Gamma^a C^{-1}\;\; \mbox{with} \ \eta^2=1.
\label{C}
\end{equation}

An appropriate resentation of the algebra (\ref{bigrad}) in which the fermionic generators satisfy the reality condition (\ref{majorana}) is
\begin{equation}
Q_{\gamma}^k= \left[ \begin{array}{cc}
0 & \delta_{\gamma}^{\alpha}\delta_j^k \\ 
-C_{\gamma \beta}u^{ki} & 0
\end{array}\right],  
\label{Q}
\end{equation}
while the bosonic generators ($B$) are accomodated in the diagonal $m\times m$ and $N \times N$ blocks,
\begin{equation}
B\sim \left[ \begin{array}{cc} (G_{(k)})_{\beta}^{\alpha} & 0 \\ 
0 & (M_{kl})_{j}^{i}
\end{array} \right].
\label{B}
\end{equation}

The anticommutator of the fermionic generators (\ref{Q}) gives
\begin{equation}
\begin{array}{l}
\left\{ Q_{\gamma}^k, Q_{\rho}^l \right\} = \\
-\left[ \begin{array}{lr}
u^{lk}(C_{\rho \beta}\delta_{\gamma}^{\alpha} + \mu C_{\gamma \beta} \delta_{\rho}^{\alpha}) & \\ 
 &C_{\gamma \rho}(u^{ki}\delta_j^l+ \lambda u^{li}\delta_j^k)
\end{array} \right]. \\
\label{QQ}
\end{array}
\end{equation}
The upper diagonal block in the r.h.s. can be expanded in a complete set of $m\times m$ matrices in the form 
\begin{equation}
C_{\rho \beta}\delta_{\gamma}^{\alpha}+ \mu C_{\gamma \beta} \delta_{\rho}^{\alpha}= A_{\gamma \rho}^{(k)}(\Gamma_{(k)})_{\beta}^{\alpha},  \label{2q}
\end{equation}
where $\Gamma_{(k)}$ stands for $\frac{1}{k!} \delta^{a_1\cdots a_k}_{b_1\cdots b_k} \Gamma^{b_1}\cdots \Gamma_{b_k}, \;\;0\leq k \leq D$  and we have used the fact that the antisymmetric product of $\Gamma$'s is a basis of $M_{m\times m}$.  

For each pair of indices ($\gamma, \rho$) the index $\alpha$ can be lowered in (\ref{2q}) multiplying by $C$.  The result is a matrix with the same symmetry as $u$,
\begin{equation}
[C\left\{ Q_{\gamma}^k, Q_{\rho}^l \right\}]_{\alpha \beta} = \mu[C\left\{ Q_{\gamma}^k, Q_{\rho}^l \right\}]_{\beta \alpha}.
\label{mu}
\end{equation}

On the other hand, the symmetry of $C\Gamma^{(r)}$ depends on the sign of $\lambda$. In fact, from (\ref{C}), it is easily seen that
\begin{eqnarray}
(C\Gamma^a)^T &=& \eta \lambda C\Gamma^a \\ 
(C\Gamma^{ab})^T &=& -\lambda C\Gamma^{ab}.
\label{rT}
\end{eqnarray}
This implies that both $C\Gamma^a$ and $C\Gamma^{ab}$ have the same symmetry --opposite to that of $C$-- only for $\eta =-1$.  Furthermore, from this and  (\ref{mu}), one concludes that the adS generators can occur in the r.h.s. of (\ref{QQ}) if and only if
\begin{equation}
\lambda \mu = \eta =-1.
\label{lm}
\end{equation}

It can be shown that for each dimension there are always two possible representations for the charge conjugation matrix ($C_1 = \Gamma_1 \Gamma_3 \cdots $, $C_2 =\Gamma_2 \Gamma_4 \cdots$).  It is easy to see that the signs of $\lambda$ fixed for all dimensions {\em except} for $D=2$ mod $4$. For  odd dimensions, $\eta =(-1)^{(D-1)/2}$, while for each even dimension $\eta= \pm 1$, with $\eta \lambda =(-1)^{[(D-2)/4]}$. 

Thus, except for $D=4k+1$, $C$ can always be chosen so that $\eta= -1$ and this uniquely fixes $\lambda$ and $\mu$.  If $\mu =1$, $u$ is a symmetric quadratic form and hence the internal group is $SO(N)$. Conversely, for $\mu =-1$ the group is $Sp(N)$. Correspondingly, if $\lambda=-1$, the only matrices that enter in the r.h.s. of (\ref{2q}) are those for which $C\Gamma$ is symmetric, which is a basis for $sp(m)$. Conversely, if $\lambda =+1$,  $C\Gamma$ is antisymmetric and the r.h.s. of (\ref{2q}) spans $so(m)$.

The semisimple algebras containing adS as a subalgebra in the bosonic sector would then be $osp(m|N)$ for $D=2,3,4$ mod 8, and $osp(N|m)$ (with even $N$) for $D=6,7,8$ mod 8. Their minimal extensions are $osp(m|1)$ and $osp(2|m)$, respectively.

The only exceptional case, for which the representation (\ref{Q}) would not be appropriate occurs for $D=5$ mod 4. In this case $\eta=+1$ and therefore the anticommutator of two $Q$'s of the form (\ref{Q}) would only contain either $\Gamma^a$ or $\Gamma^{ab}$, but not both and therefore it would not correspond to a supersymmetric extension of adS.  In this case, the analysis is best carried out using complex Dirac spinors.  One can repeat the previous construction but taking the fermionic generators of the superalgebra in the complex representation 
\begin{eqnarray}
\bar{Q}_{\gamma }^{l}&=&\left[ \begin{array}{lr}
0 &\delta_{\gamma}^{\alpha}\delta_j^l \\ 
0 & 0
\end{array} \right  ], \\ 
Q_{\rho k}        &=&\left[ \begin{array}{lr}
0 & 0 \\ 
-G_{\rho \beta}\delta_k^i & 0
\end{array} \right],
\label{repU}
\end{eqnarray}
where $G$ is the Dirac conjugation matrix (for minkowskian signature, $G=\Gamma^0$). Now the anticommutator of $Q$'s does not have a definite symmetry. In this case, the charge conjugation matrix can be chosen to be antihermitian and the upper block of $\{ \bar{Q}, Q \}$ closes on the unitary algebra $u(m-2,2)$, while the lower block spans $u(N)$. Thus, for $D=5$ mod $4$, the superalgebra is $u(m-2,2|N)$, whose minimal extension is $su(m-2,2|1)$.

The previous discussion can be summarized in the following table\footnote{A special case occurs for $D=3$, where we have defined the supergroup as $osp(2|1)$ only for completeness of the table. This group does not contain the generator of adS and, stricktly speaking, one should have written $osp(2|1)\otimes sp(2)$ instead.}:

\[\begin{tabular}{|c|l|l|l|l|}
\hline
{\bf D}& $\lambda$ & $\eta$  & $\mu$   &{\bf Superalgebras} \\ \hline
  2    & -1\ [+1]  & -1\ [+1]& +1\ [-1]& $osp(2|N)$  [$osp(N|2)$] \\ 
  3    & -1        & -1      & +1      & $osp(2|N)$  \\ 
  4    & -1        & -1\ [+1]& +1      & $osp(4|N)$  \\ 
  5    & -1        & +1      & +1      & $u(4|N)$    \\ 
  6    & +1\ [-1]  & -1\ [+1]& -1\ [+1]& $osp(N|8)$  [$osp(8|N)$]\\ 
  7    & +1        & -1      & -1      & $osp(N|8)$  \\ 
  8    & +1        & -1\ [+1]& -1      & $osp(N|16)$ \\ 
  9    & +1        & +1      & -1      & $u(16|N)$   \\ 
  10   & -1\ [+1]  & -1\ [+1]& +1\ [-1]& $osp(32|N)$  [$osp(N|32)$]\\ 
  11   & -1        & -1      & +1      & $osp(32|N)$ \\ 
  12   & -1        & -1\ [+1]& +1      & $osp(64|N)$ \\ 
  .    &  .        &    .    &  .      & . \\ 
  .    &  .        &    .    &  .      & . \\ 
  .    &  .        &    .    &  .      & . \\ \hline
\end{tabular} \]
We indicate in brackets the alternative choices that satisfy $\lambda \mu =-1$ but $\eta=+1$. The corresponding supersymmetric theories do not have both the generator $J^a$ in the bosonic sector and therefore, the contact with gravity would be less transparent. Note that these ``pseudo supergravities" exist for all even dimensions. Their algebras are different from those of the ``real supergravities" for $D=2$ mod 4, but are the same for $D=4$ mod 4.

For $D\neq 5$ mod 4, the bosonic generators take the form
\begin{equation}
J_{(k)} = \left[ \begin{array}{cc} \frac{1}{2}(\Gamma_{(k)})_{\beta}^{\alpha} & 0 \\ 
0 & 0
\end{array} \right].
\label{j}
\end{equation}
\begin{equation}
M^{kl} = \left[ \begin{array}{cc} 0 & 0 \\ 
0 & (m^{kl})_j^i
\end{array} \right].
\label{m}
\end{equation}
with $(m^{kl})_j^i = u^{ki}\delta^l_j + \lambda u^{li}\delta^k_j $. The superalgebra now reads
\begin{eqnarray}
[ J^{(p)}, J^{(q)} ]     & = & f^{(p)(q)}_{(r)} J^{(r)} \\ \nonumber
[ J^{(p)}, M^{jl}  ]     & = & 0 \\ \nonumber
[ J^{(p)},Q^l_{\alpha} ] & = & \frac{1}{2}\Gamma^{(p)})^{\beta}_{\alpha}
Q^l_{\beta} \\ \nonumber
[ M^{ij}, M^{kl}  ]      & = & f^{ijkl}_{hm} M^{hm}  \\ \nonumber
[ M^{jk}, Q^l_{\alpha} ] & = & (m^{jk})^l_i Q^i_{\alpha} \\ \nonumber
\{ Q^k_{\gamma}, Q^l_{\rho}  \} & = &\frac{4}{m}u^{lk} (C\Gamma^{(p)})_{\gamma \rho}J_{(p)} - C_{\gamma \rho} M^{kl}.
\end{eqnarray}

\subsection{Equations of Motion}

Let $\Omega$ be a $2n$-dimensional manifold with boundary $\partial \Omega
=\Sigma$. Then, by Stokes' theorem the action can be written as

\begin{equation}
I_{C-S}=\int_{\Sigma }L=\int_{\Omega }dL=\int_{\Omega }\left\langle\mbox{{\bf F}}^n\right\rangle.
\end{equation}

Varying $I$ with respect to the connection, using $\delta \mbox{{\bf F}} =\nabla \delta${\bf A} and the Bianchi identity $\nabla${\bf F}$=0$, we obtain

\begin{equation}
\delta I_{C-S}=n\int_{\Omega }\left\langle\nabla (\delta \mbox{{\bf A}})\mbox{\tiny$ \wedge$} \mbox{{\bf F}}^{n-1}\right\rangle =n\int_{\Omega}d\left\langle\delta\mbox{{\bf A}}\mbox{\tiny $\wedge$} \mbox{{\bf F}}^{n-1}\right\rangle.
\end{equation}

Using Stokes' theorem again, 
\begin{equation}
\delta I_{C-S}=n\int_{\Sigma }\delta \mbox{{\bf A}}^B \mbox{\tiny $\wedge$}\left\langle G_B \mbox{{\bf F}}^{n-1}\right\rangle.
\label{deltaI}
\end{equation}

Thus, the equations of motion are
\begin{equation}
\left\langle\mbox{{\bf F}}^{n-1}G_{B}\right\rangle =0.
\label{eccs}
\end{equation}

These equations provide a representation of the supergravity algebra. Suppose the superalgebra has generators $G_{A}=\left\{ B_{a};Q_{\alpha }\right\} $,
with a graded commutator algebra of the form

\begin{equation}
\begin{array}{ccc}
\left[ B_{a},B_{b}\right] & = & C_{ab}^{c}B_{c} \\ 
&  &  \\ 
\left[ B_{a},Q_{\beta }\right] & = & C_{a\beta }^{\gamma }Q_{\gamma } \\ 
&  &  \\ 
\left\{ Q_{\alpha },Q_{\beta }\right\} & = & C_{\alpha \beta }^{c}B_{c}
\end{array}
\label{Thomas}
\end{equation}

The connection associated to this algebra is {\bf A}$=W^aB_a +\Psi^{\alpha} Q_{\alpha} = W+\Psi$. Calling $D$ the covariant derivative restricted to the bosonic subgroup generated by $B_a$. By virtue of the Bianchi identity $\nabla${\bf F}$=0$, we obtain

\begin{equation}
D \mbox{{\bf F}}=\left[ \mbox{{\bf F}},\Psi \right] .  \label{DF}
\end{equation}

Varying the action with respect to the connection $\delta${\bf A}$=\delta W+ \delta \Psi$, yields the field equations

\begin{equation}
\begin{array}{ccc}
\delta W: &  & \left\langle\mbox{{\bf F}}^{n-1}B_a \right\rangle =0 \\ 
\delta \Psi : &  & \left\langle\mbox{{\bf F}}^{n-1}Q_{\alpha }\right\rangle =0
\end{array}
\end{equation}

Acting with $D$ on the fermionic equation gives
\begin{equation}
D\left\langle\mbox{{\bf F}}^{n-1}Q_{\alpha }\right\rangle =\left\langle D(\mbox{{\bf F}}^{n-1})Q_{\alpha }\right\rangle = \left\langle\mbox{{\bf F}}^{n-1}\left\{ \Psi,Q_{\alpha }\right\} \right\rangle=0,
\end{equation}
where we have used (\ref{DF}) and the symmetry property of the bracket. Since  $\left\{ \Psi,Q_{\beta }\right\} =\Psi^{\alpha }\left\{ Q_{\alpha}, Q_{\beta}\right\} =\Psi ^{\alpha }C_{\alpha \beta }^cB_c$, the consistency condition is

\begin{equation}
\left\langle\mbox{{\bf F}}^{n-1}B_c\right\rangle C_{\alpha \beta }^c \Psi^{\alpha}=0.
\end{equation}

Supposing that $\Psi^{\alpha }$ is an arbitrary solution of the fermionic field equations, we conclude that

\begin{equation}
\left\langle\mbox{{\bf F}}^{n-1}B_c\right\rangle C_{\alpha \beta}^c=0.  \label{cons}
\end{equation}

Thus, the consistency conditions associated with the fermionic equations do not produce additional restrictions on the theory. They are a subset of the bosonic field equations.  If the structure constants $C_{\alpha \beta }^{c}$ were nonzero, the complete set of bosonic equations would enter in Eq. (\ref{cons}). The preceding analysis of the integrability conditions holds equally for Chern-Simons supergravities as for a standard supersymmetric theory.

\subsection{Contraction of Super-$adS$}

 In \cite{btrz}, a family of CS theories were constructed for supersymmetric extensions of the Poincar\'e algebra of the form (\ref{supertrans}). 
\begin{equation}
\{Q^{\alpha },\bar{Q}_{\beta }\}=-i(\Gamma ^{a})_{\beta }^{\alpha}P_{a}
-i(\Gamma^{abcde})_{\beta }^{\alpha }Z_{abcde}. 
\label{supertrans'}
\end{equation}

It is possible to obtain this algebra by an appropriate contraction of the $N=2$ superextension of an adS algebra, in the limit of vanishing cosmological constant. Let us define 
\begin{equation}
T_{M}=\left( \begin{array}{c}
J_{a} \\ 
J_{ab} \\ 
J_{abcde} \\ 
Z_{(m)} \\ 
Q \\ 
\bar{Q} \\ 
M_{ij}
\end{array} \right), 
\label{Tm}
\end{equation}
where $Z_{(m)}$ stands for all the $J_{(m)}$'s with $m\neq 1,2,5$, and $Q=Q^{1}+iQ^{2}$, $\bar{Q} =Q^{1}-iQ^{2}$.  Consider now the contraction  of the form $T_{M}^{\prime}= U_{M}^{K}(\epsilon )T_{K}$, where
\begin{equation}
U_{M}^{K}(\epsilon )=\left[ \begin{array}{ccccccc}
f_{p}(\epsilon ) &  &  &  &  &  &  \\ 
& 1 &  &  &  &  &  \\ 
&  & f_{k}(\epsilon ) &  &  &  &  \\ 
&  &  & 1 &  &  &  \\ 
&  &  &  & f_{q}(\sqrt{\epsilon}) &  &  \\ 
&  &  &  &  & f_{q}(\sqrt{\epsilon}) &  \\ 
&  &  &  &  &  & 1
\end{array} \right]
\label{contractor}
\end{equation}
Here $f_x(\alpha )=\alpha ((1-x)\alpha +x)$, so that $f_x(1)=1$ and  $\lim_{\alpha \rightarrow 0} f_x(\alpha )\sim x\alpha$.  Thus, it can be easily checked that for $\epsilon \rightarrow 0$ the generators $T_M^{\prime}$ satisfy a closed superalgebra, where the only nonvanishing anticommutator is of the form (\ref{supertrans'}),
\begin{equation}
\{ Q^{\prime}, \bar{Q}^{\prime} \} =\frac{4q^2}{m}(\frac{\Gamma^a}{p}
J^{\prime}_a+ \frac{\Gamma^{abcde}}{k}J^{\prime}_{abcde}). \label{supertrans''}
\end{equation}

The dimensionless parameters $q,p,k$ can be eliminated by rescaling the gauge fields. They can also be interpreted as coupling constants in the resulting gauge theory. 

It is remarkable that for $D=3,5,7,11$, it is always possible to contract the algebra so that the internal gauge group {\bf M}, generated by $M_{ij}$ decouples from the Poincar\'e generators. If we call {\bf G} the group generated by (\ref{supertrans'}), the contraction produces a semidirect product {\bf G}$\odot ${\bf M,}, where the only nonvanishing commutator is 
\begin{equation}
\left[ Q_{k}^{^{\prime }\alpha },M_{ij}\right] =\delta _{\beta }^{\alpha} c_{ijk}^{l}Q_{l}^{^{\prime}\beta}.
\end{equation}

For $D=9$ and $D>11$, on the other hand, the contraction contains {\bf G}$\odot ${\bf M} as a proper subgroup, since the generators $Z_{(m)}$ carry vector indices, and therefore cannot be contained in {\bf M} \footnote{Note that, in this sense, $D=11$ is an upper bound from a purely geometrical argument.}. It is then reasonable to expect that, in analogy with what happens in three dimensional gravity, the CS Poincar\'e supergravity theories of \cite{btrz} could be obtained from a contraction of the CS adS supergravities.


\end{document}